\documentclass[fleqn,usenatbib]{mnras}

\usepackage{newtxtext,newtxmath}
\usepackage[T1]{fontenc}
\usepackage{ae,aecompl}
\usepackage{graphicx}
\usepackage{amsmath}
\usepackage{amssymb}
\usepackage{lipsum}

\usepackage{multirow}

\newcommand\lsim{\mathrel{\rlap{\lower4pt\hbox{\hskip1pt$\sim$}}
        \raise1pt\hbox{$<$}}}
\newcommand\gsim{\mathrel{\rlap{\lower4pt\hbox{\hskip1pt$\sim$}}
        \raise1pt\hbox{$>$}}}

\title[Correlation between Optical and UV Variability of Quasars]{Correlation between Optical and UV Variability of Quasars}

\author[C.~Xin et al]{Chengcheng~Xin,$^{1}$
Maria~Charisi,$^{2}$\thanks{E-mail: mcharisi@caltech.edu}
Zolt{\'{a}}n~Haiman,$^{1}$ David Schiminovich$^{1}$
\\
$^{1}$Department of Astronomy, Columbia University, New York, NY, 10027\\
$^{2}$TAPIR, California Institute of Technology, Pasadena, CA, 91125}

\date{Accepted XXX. Received YYY; in original form ZZZ}

\pubyear{2019}

\begin{document}
\label{firstpage}
\pagerange{\pageref{firstpage}--\pageref{lastpage}}
\maketitle

\begin{abstract}
The variability of quasars across multiple wavelengths is a useful
probe of physical conditions in active galactic nuclei. In particular, variable accretion rates, instabilities, and reverberation effects in the accretion disk of a supermassive black hole (SMBH) are expected to produce correlated flux variations in UV and optical bands. Recent work has further argued that binary quasars should exhibit strongly correlated UV and optical periodicities. Strong UV-optical correlations have indeed
been established in small samples of ($N\lsim 30$) quasars with well-sampled light curves, and have extended the "bluer-when-brighter" trend previously found within the optical bands. Here we further test the nature of quasar variability by examining the observed-frame UV-optical correlations in a large sample of 1,315 bright quasars with overlapping UV and optical light curves for the Galaxy Evolution Explorer ({\it GALEX}) and the Catalina Real-time Transient Survey ({\it CRTS}), respectively. We find that strong correlations exist in this much larger sample, but we rule out, at $\sim$95\% confidence, the simple hypothesis that the intrinsic UV and optical variations of all quasars are fully correlated. Our results therefore imply the existence of physical mechanism(s) that can generate uncorrelated optical and UV flux variations.
\end{abstract}

\begin{keywords}
quasars: general  -- galaxies: active -- stars: variables: others
\end{keywords}

\section{Introduction}\label{sec:intro}

Quasars are thought to be powered by gas accreted onto supermassive black holes
\citep{Lynden1969}. 
The black hole is fed by a geometrically thin accretion disk, whose thermal emission is stratified radially, with higher-energy radiation arising from the hotter inner regions (e.g., \citealt{Frank2002}).

A prominent feature of quasars is their
variability across multiple wavelengths.  Quasars appear to vary on time-scales from minutes to years and their variability has been studied extensively.
This is especially true in optical bands, where time-domain surveys have provided large samples of quasar light curves. The variability is stochastic, and
well described by random Gaussian fluctuations, whose auto-correlation
function obeys the so-called ``damped random walk'' (DRW) model
(e.g., \citealt{Macleod2010,Kelly2009}; 
but see \citealt{Smith+2018} for a discussion on discrepancies from this model).  
Similar behavior is also seen in the ultraviolet (UV) band, with generally larger variability amplitudes \citep{Welsh2011,Gezari2013,Zhu2016}.

The origin of the stochastic variability is currently poorly understood, but in general, fluctuations in the accretion rate \citep{Pereyra2006,Schmidt2012a,Li2008,Ruan2014}, instabilities or
inhomogeneities in the accretion disk \citep[e.g.][]{Dexter2011}, and reprocessing of
variable higher-energy (X-ray) emission from a hot corona within a few gravitational
radii of the central SMBH \citep{Krolik1991,George1991} could all cause variability in optical and UV bands.

All of the above mechanisms
are expected to induce variability that is correlated across wavelengths.  Indeed, cross-correlations within the optical bands are well established, with a clear "bluer-when-brighter" trend \citep{VandenBerk2004,Wilhite2005,Schmidt2012a}. These inter-band correlations
should be especially useful to constrain physical models, since in the
radially stratified disks, the UV and optical emission regions are
expected to be spatially well separated and therefore need not necessarily co-vary (e.g., if there are localized temperature fluctuations in an inhomogeneous disc; \citealt{Dexter2011}).
For example, inter-band cross-correlations have been used to test whether variability
can be fully explained by changes in the accretion rate of a quasi-steady disc,
with early work finding consistency \citep{Pereyra2006}, but recent studies finding that accretion rate variations alone cannot fully explain the observed color variability~\citep{Schmidt2012a,Ruan2014,Kokubo2014}.  Likewise, localized temperature fluctuations appear not to be the main driver of optical color variability~\citep{Ruan2014, Kokubo2015}. 

Cross-correlations have also been established between the UV and optical bands (\citealt{Hung2016,Edelson2019}). Because UV data are typically sparse, 
previous studies of the optical-to-UV cross-correlations have been performed only on small samples.  For example, \citealt{Edelson2019} relied on
well-sampled light curves, obtained via intense monitoring of four AGN
in optical and UV (as well as X-ray) bands with {\it Swift}. They find strong
UV/optical cross-correlations, with time-lags (on the order of a $\tau\sim$ day) whose value is larger,
but the scaling consistent with $\tau \propto \lambda^{4/3}$ expected in a radially stratified and centrally illuminated thin disk.
Similar results were found by \citet{Fausnaugh+2016} for the Seyfert galaxy NGC 5548 with an active nucleus, using ground-based optical monitoring, combined with UV data from the {\it Hubble Space Telescope} and {\it Swift}. \citet{Hung2016} analysed the wavelength-dependent variability of 23 AGN with large optical variability in the Panoramic Survey Telescope and Rapid Response System (Pan-STARRS), 
combined with two epochs of UV data, separated by approximately a year, in the {\it GALEX} Time Domain Survey. They found that the UV/optical correlations for 17 out of the 23 quasars are consistent with the variable accretion-rate disk model.

Additionally, recent work has identified possibly significant optical periodicity in a small fraction of quasars \citep{Graham2015b,Charisi2016,Liu2018}. These quasars have been proposed to host binary SMBHs, whose orbital motion is expected to induce periodic variability due to hydrodynamical modulations in the accretion rate (e.g., \citealt{Hayasaki2008,MM08,Roedig+11,Shi2012,DOrazio2013,Farris2014,Munoz2019}),
as well as due to relativistic Doppler boost of the emission from gas bound to individual SMBHs \citep{Dorazio2015Nature, Charisi2018, Xin2019}. The latter effect induces strongly correlated variability in optical and UV bands with amplitudes that depend on the spectral curvature. Given the limited quality of currently available UV light curves, Doppler-boost variability can be confused with the multi-wavelength variability of aperiodic single-SMBH quasars \citep{Charisi2018}.

In this paper, we complement previous work by examining the UV {\it vs.} optical cross-correlations in a large sample of quasars.  Our
motivation is to assess whether tight correlations are generic and
exist in all quasars, or if there are examples of uncorrelated, or
only partially correlated optical and UV flux variations. 
In order to
do this, we created a sample of $>1,000$ sources, which have
adequately sampled and overlapping UV and optical time-series,
from {\it GALEX} and {\it CRTS}, respectively.  By analysing the covariance between the pairs
of light curves, we find that strong correlations exist in this larger
sample, but we rule out, at $\sim$95\% confidence, the simple
hypothesis that the intrinsic UV and optical variations of all quasars
are fully correlated.  Our results therefore imply the existence of
physical mechanism(s) that can generate unrelated optical and UV flux
variations.

The rest of this paper is organised as follows.
In \S~\ref{sec:sample_and_data}, we describe our methodology,
including the UV and optical quasar catalogs we used
(\S~\ref{sec:quasar_catalogs}), the construction of our joint
UV+optical sample of 1,315 sources (\S~\ref{sec:sample_selection}), and
our analysis of the light curves, including the computation of the
cross-covariance between the noisy and sparsely and heterogeneously sampled
pairs of time-series (\S~\ref{sec:data_analysis}).
In \S~\ref{sec:results} we present our findings, in the form of the
distribution of the cross-correlation coefficients found for the 1,315
quasars. These distributions are determined both by intrinsic
correlations, as well as the data quality; we compare them
with the distributions predicted under different assumptions of the
underlying {\em intrinsic} correlations.
In \S~\ref{sec:discussion} we discuss these results, including the
impact of the poor sampling, limited baselines, and photometric noise,
as well as the possibility of time-lags and partial correlations.
Finally, we summarize our findings and conclusions in
\S~\ref{sec:conclusion}.

\section{Sample Selection and Data Analysis}\label{sec:sample_and_data}

\subsection{UV and Optical Data}
\label{sec:quasar_catalogs}
In order to study the observed-frame correlation between UV and optical variability in a large sample of quasars, we extract UV and optical light curves from {\it GALEX} and {\it CRTS}, respectively. The combination of these two surveys is optimal, because both are all-sky surveys, covering large samples of quasars, the light curves of which have significant temporal overlap.\footnote{Note that {\it GALEX} is the only wide and deep time-domain survey in UV, whereas in the optical, Pan-STARRS and the Palomar Transient Factory (PTF) are less ideal for this study, because they began operations after {\it CRTS}, and their overlap with {\it GALEX} is more limited.}

In particular, {\it GALEX} \citep{Martin2005}
was the first all-sky survey in the UV band. {\it GALEX} operated from 2003 to 2012 and provided simultaneous photometric measurements in two UV filters, in the far-UV ($1350-1750$ \AA) and near-UV ($1750-2750$ \AA) bands (hereafter FUV and NUV, respectively). 
It performed three main surveys: (i) an all-sky imaging survey (40,000\,deg$^2$) with a limiting magnitude of $m_{AB}\sim20.5$, (ii) a medium imaging survey (1000\,deg$^2$) with limiting magnitude $m_{AB}\sim23$, and (iii) a deep imaging survey (100\,deg$^2$) with limiting magnitude $m_{AB}\sim25$.
In addition, from 2008 to 2011, {\it GALEX} performed a time-domain survey (40\,deg$^2$) with a 2-day cadence  \citep{Gezari2013}.

{\it CRTS} \citep{Drake2009} is an ongoing time-domain survey, which covers the majority of the sky from declinations of $-75^{\circ}$ to $65^{\circ}$, with the exception of the galactic plane, in unfiltered optical light, broadly calibrated in the Johnson $V$-band. It began operations in 2005
and the most recent data release (Data Release 2), which we use, extends to 2014.\footnote{Even though {\it CRTS} continues operations, the most recent data are not publicly available.}
{\it CRTS} covers up to $\sim$2500 deg$^2$ per night
and each visit consists of four exposures, separated by 10 minutes.

\subsection{Sample Selection}\label{sec:sample_selection}

Our basic approach is to study as large a population of quasars as possible. For this, we first select the sample based on the UV light curves, and after necessary quality cuts, we cross-correlate with {\it CRTS}. This strategy maximizes the sample of quasars with suitable data, since it optimizes the selection for the UV sample, which typically has lower quality data. 
We note that even with this strategy, it is still necessary to compromise on the quality of the individual light curves, in order to maintain a large sample.

Our starting sample is the Half Million Quasars catalogue (HMQ; version 4.4; \citealt{Flesch2015}), which consists of 424,748 spectroscopically confirmed quasars. 
We extracted sources within 5\,arcsec from the input position using the final {\it GALEX} data release (DR6/7).\footnote{\url{https://asd.gsfc.nasa.gov/archive/galex/}} Of the quasars in the HMQ catalogue, 159,750 have at least one observation in the NUV band.
The extracted sample contains a broad population of quasars at various redshifts (up to $z\approx5$), and NUV magnitudes ($16\lsim m_{\rm AB} (NUV) \lsim 24$), as shown by the dark blue points and curves in Fig.~\ref{fig:redshift-mag} below.

In Fig.~\ref{fig:visit-bl}, we show the number of observations versus the baseline of the light curves for the entire {\it GALEX} sample
in both the NUV and FUV bands.
The typical light curves from {\it GALEX} are sparse with only a few epochs, but there is a large variety in terms of number of observations and temporal baselines ranging from a few days up to 10 years, similar to the findings in \citet{Welsh2011}.
It is also clear that the selected quasars have fewer data points in FUV compared to the NUV band, as expected due to the fact that the FUV detector was operational over a shorter time span.
For this reason, in the remainder of this paper, we focus only on the NUV light curves.

\begin{figure}
 \includegraphics[width=\columnwidth]{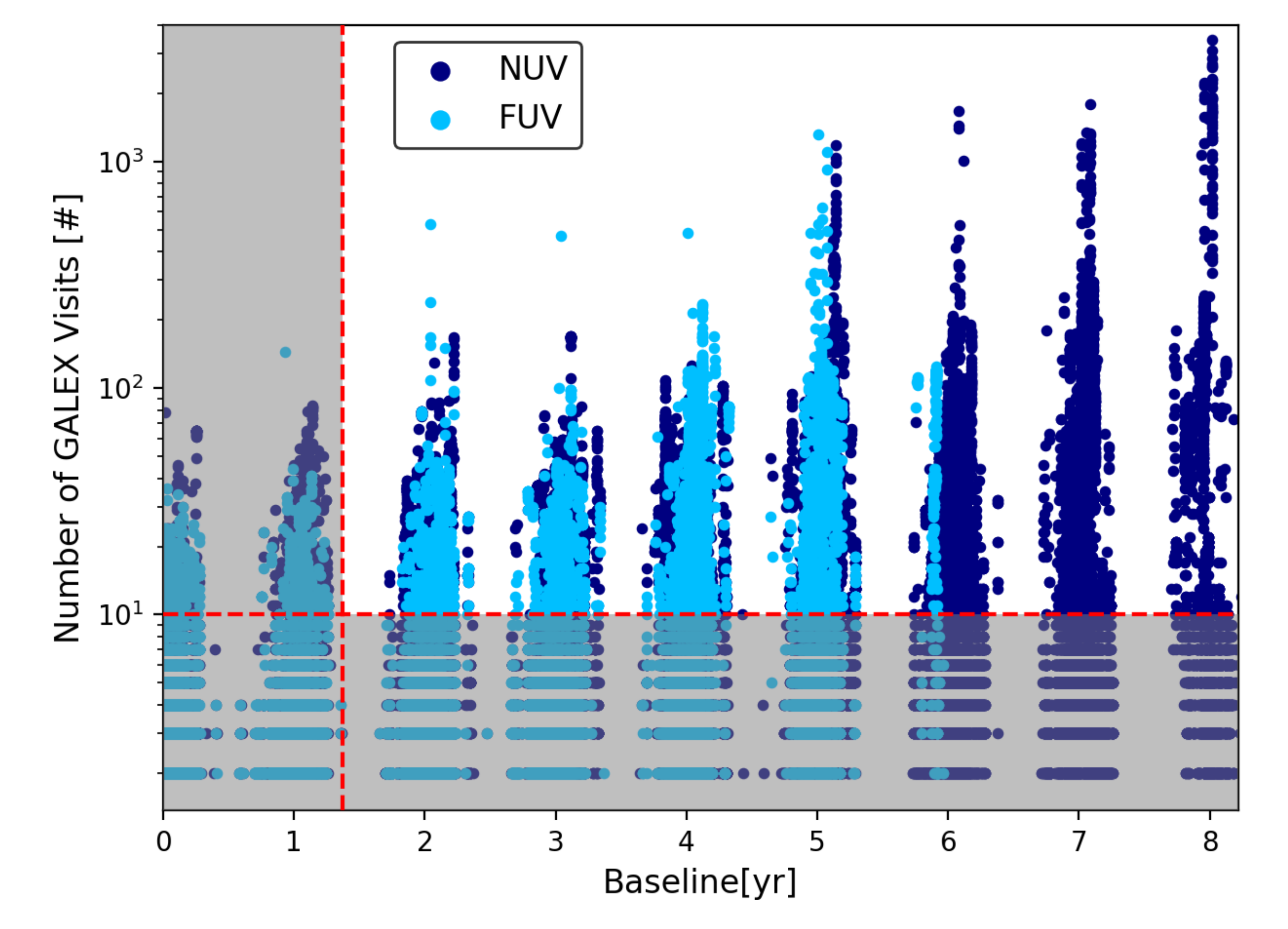}
 \caption{The number of {\it GALEX} visits
 are shown against the total baseline for the entire NUV and FUV samples, in dark and light blue, respectively. The red dashed lines denote necessary quality cuts on the light curves (i.e. at least 10 visits and baselines of 500\,d). These cuts exclude the sources in the shaded area from our final sample.}
 \label{fig:visit-bl}
\end{figure}
 
\begin{figure}
 \includegraphics[width=\columnwidth]{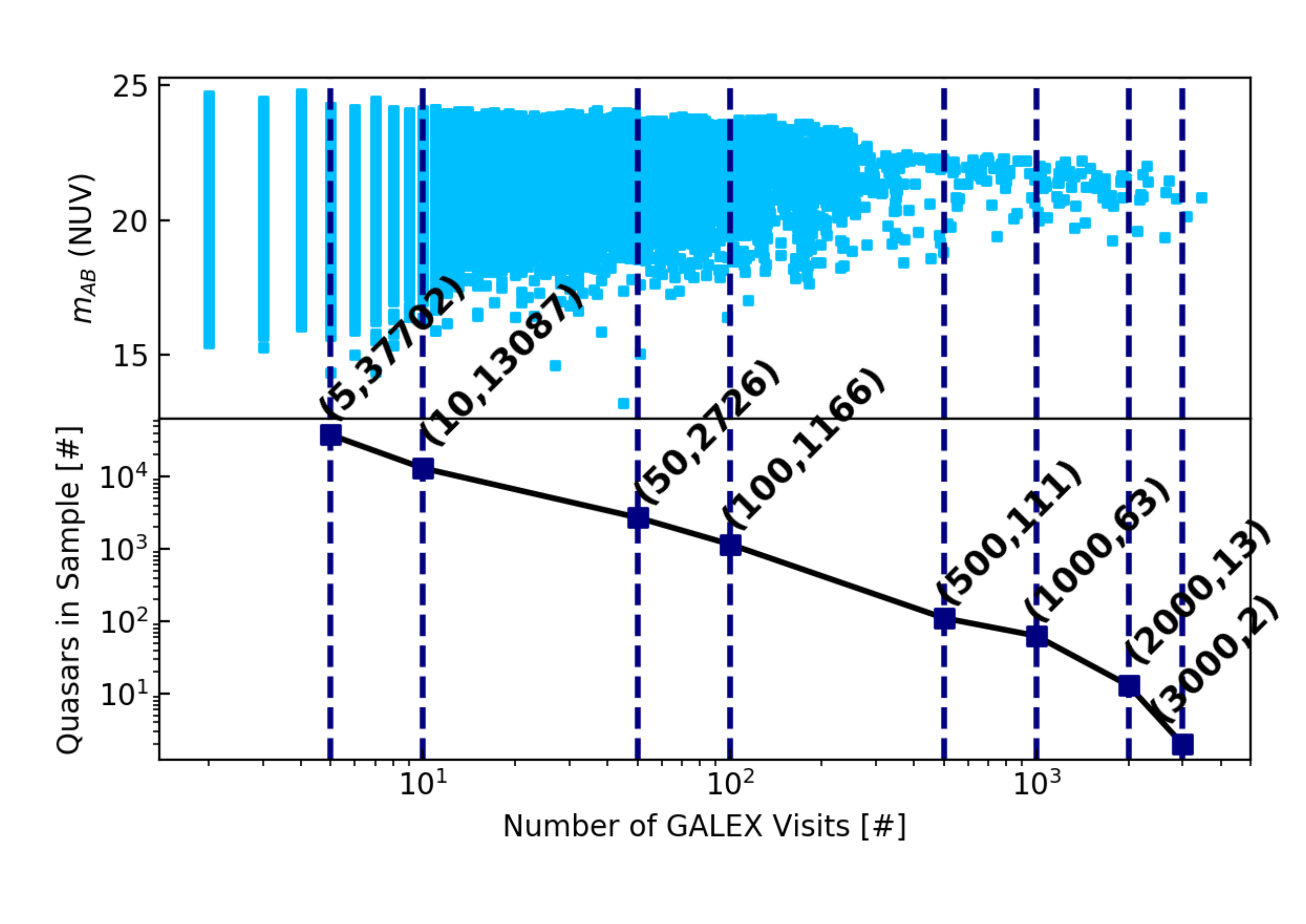}
 \caption{The top panel shows the mean NUV magnitude {\em vs.} the number of {\it GALEX} visits for each source. The dashed lines mark cuts corresponding to various different minimum number of visits. The bottom panel shows the cuts and the number of quasars left in the sample after each cut, which are indicated with two numbers in the parentheses.}
 \label{fig:visits-mean}
\end{figure}

In order to avoid possible biases associated with too few observations, we select light curves based on the number of data points. In Fig.~\ref{fig:visits-mean} (bottom panel), we show the number of quasars remaining in the sample as a function of the minimum allowed number of observations. As expected, requiring more visits results in a declining number of available quasars. We set the minimum required number of {\it GALEX} visits to 10, which ensures relatively more frequently sampled sources, while maintaining a large sample of 13,087 quasars. In Fig.~\ref{fig:visits-mean} (top panel), we show the mean NUV magnitude versus the number of {\it GALEX} visits; the NUV catalog (both initially and after the aforementioned quality cut) consists of quasars with a wide range of mean photometric magnitudes.

Among these 13,087 {\it GALEX} NUV sources, we exclude light curves with baselines shorter than $\sim$500 days. We have found it necessary to exclude these light curves, because for the cross-correlation analysis, we need at least three distinct well-separated epochs.\footnote{In the UV light curves, we define an epoch as a cluster of time-ordered points separated with gaps smaller than 30 days.  While somewhat ad-hoc, we have found that in practice, this definition identifies the discrete clusters of data-points (e.g. $\approx$10 such clusters are seen among the gray points Fig.~\ref{fig:method-obs}).}
Light curves spanning less than $\sim$500 days are typically observed for only two epochs, as can be inferred from the sampling pattern in Fig.~\ref{fig:visit-bl}.   In addition, the typical DRW timescale $\tau$ is a few 100 days. As a result, in order to fairly sample the light-curves, the overlap should cover longer periods for an unbiased measurement of cross-correlations.
This additional requirement resulted in a relatively minor cut, eliminating only 900 of the 13,087 quasars.

The remaining 12,187 quasars in the {\it GALEX} UV catalog are then cross-matched with their optical counterparts in the {\it CRTS} catalog.\footnote{\url{http://nesssi.cacr.caltech.edu/DataRelease/}} We extract light curves within 3\,arcsec from the HMQ catalogue coordinates, which returns a total of 2,840 quasars.

\begin{figure}
 \includegraphics[width=\columnwidth]{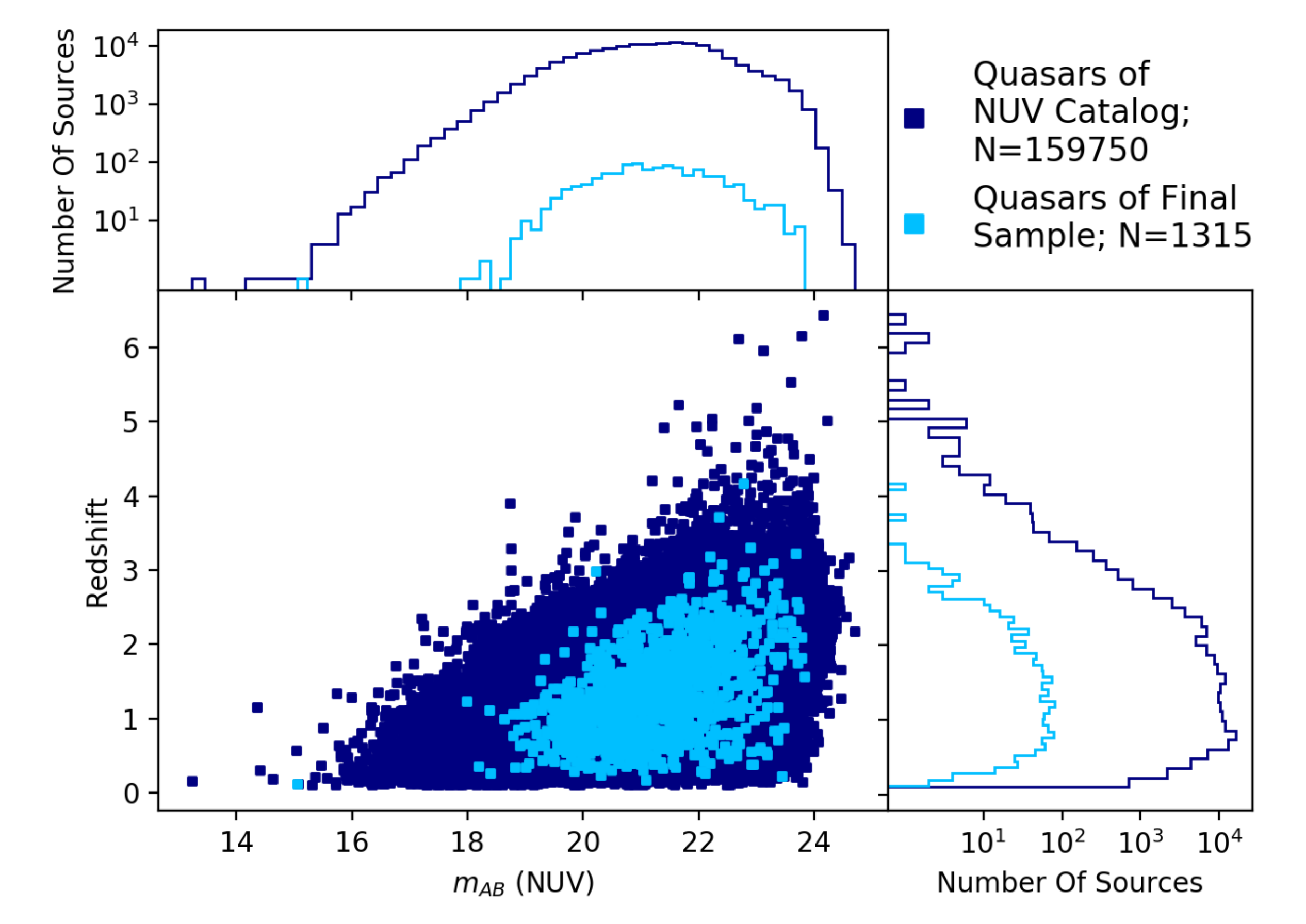}
 \caption{The distribution of redshifts and mean NUV magnitudes of 159,750 quasars extracted from {\it GALEX} (dark blue) and the final sample of 1,315 quasars used in our analysis (light blue).}
 \label{fig:redshift-mag}
\end{figure}

For the cross-correlation analysis, we need overlapping optical and UV data, which imposes further cuts on the sample. 
For this, we identified the temporal overlap between the UV and optical light curves of each source, and excluded sources whose NUV light curve did not cover at least three distinct epochs within the optical baseline.
Finally, we then excluded sources with sparse optical data and with unrealistically high virial black hole mass estimates (see \S~\ref{sec:modeling-simulation} below).  Overall, these selections result in 1,315 high-quality (according to our criteria) pairs of light curves.

\begin{figure}
 \includegraphics[width=\columnwidth]{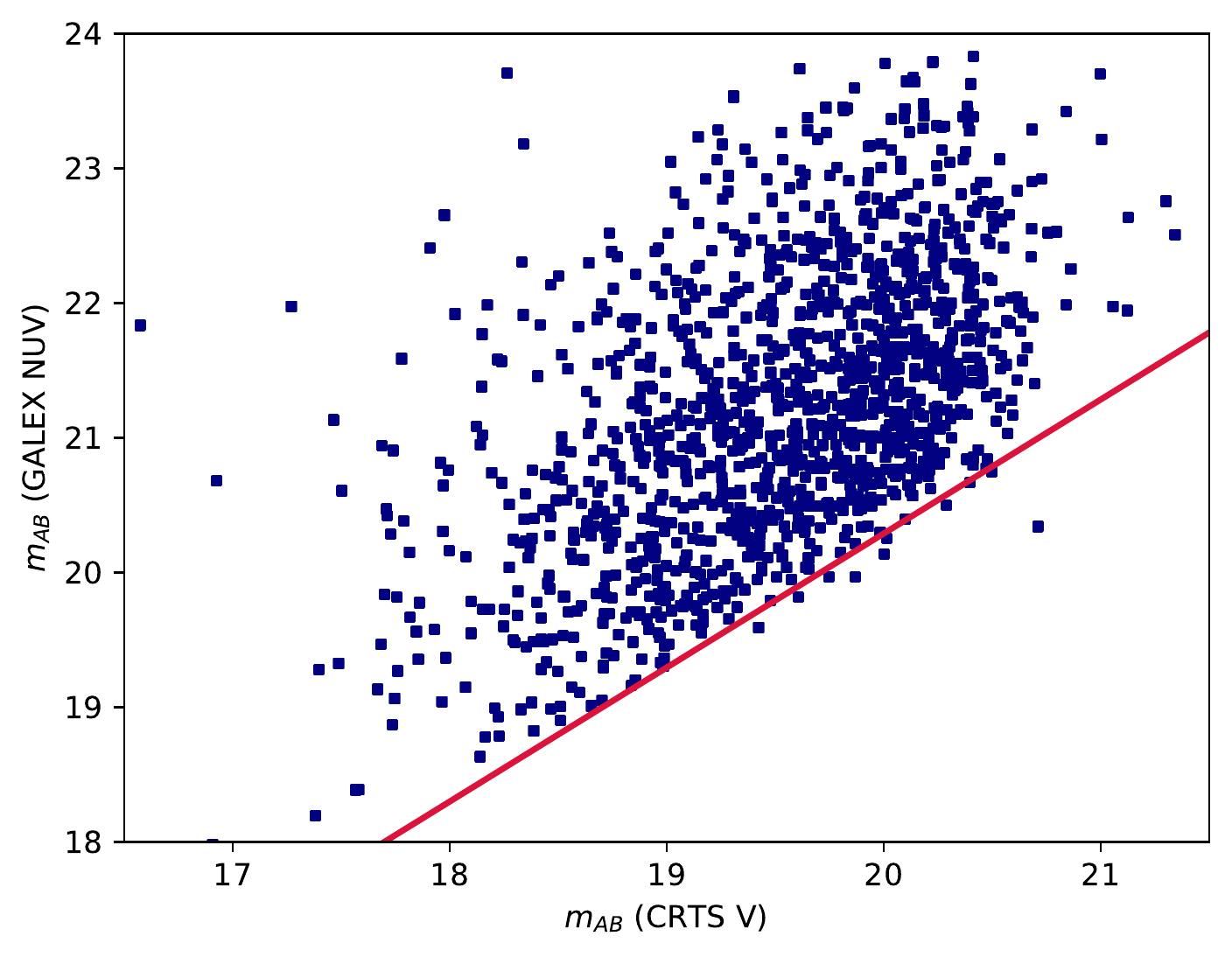}
 \caption{Mean photometric optical and NUV magnitudes of the 1,315 sources in our final sample.  For reference, the diagonal red line shows when the two magnitudes are equal.}
 \label{fig:mean_mag}
\end{figure}

In Fig.~\ref{fig:redshift-mag}, we show the distribution of redshift and NUV magnitude for the initial sample extracted from {\it GALEX} along with the final sample of 1,315 quasars.
The properties of our final sample resemble those of the input parent sample, without introducing significant biases. Specifically, our final sample covers broad ranges of redshifts and magnitudes, similar to those in the original {\it GALEX} and {\it CRTS} catalogs. 
Fig.~\ref{fig:redshift-mag} shows that the redshift distribution of our sample preserves the peak at $z\approx1$, although it also shows that our selection removes sources with higher redshifts, which tend to be dim with sparse light curves. The magnitude distribution is similar to the initial sample, except that sources at the rare faint and bright tails of the distribution are missing from the final sample. For the final sample, we also illustrate the mean NUV magnitude versus the mean $V$-band magnitude,  in Fig.~\ref{fig:mean_mag}.
The magnitudes in the two bands are tightly correlated, and the optical magnitude is typically brighter than the NUV, as expected from the typical spectral energy distribution of quasars.

\subsection{Data Analysis}\label{sec:data_analysis}
\subsubsection{Observed Data} \label{sec:observed_data}

In order to characterize the correlations between optical and UV variability in the sample as a whole, we first quantify the cross-correlation for each of the 1,315 sources individually. We first pre-processed the light curves (optical and NUV) removing outliers and binning in one day intervals, following the steps in \citet{Charisi2016}. Note that the short-term variability is not significant for the cross-correlation analysis and thus binning does not affect our results.

Calculating cross-correlations with our data is challenging, because the observations in the two bands are not taken simultaneously. To address this, we interpolated the optical light curves with a polynomial. We chose to interpolate the optical light curves, because the quality of the data is significantly higher than in the NUV (i.e. the optical data typically have more epochs, smaller gaps and lower photometric uncertainty). We found the best-fit polynomial for each binned optical light curve by varying the polynomial order from $n=6$ to $n=30$, and choosing the smallest $n$ that produces a reduced $\chi^2$ close to unity. This process ensures reasonable fits, while avoiding over-fitting.

\begin{figure}
 \includegraphics[width=\columnwidth]{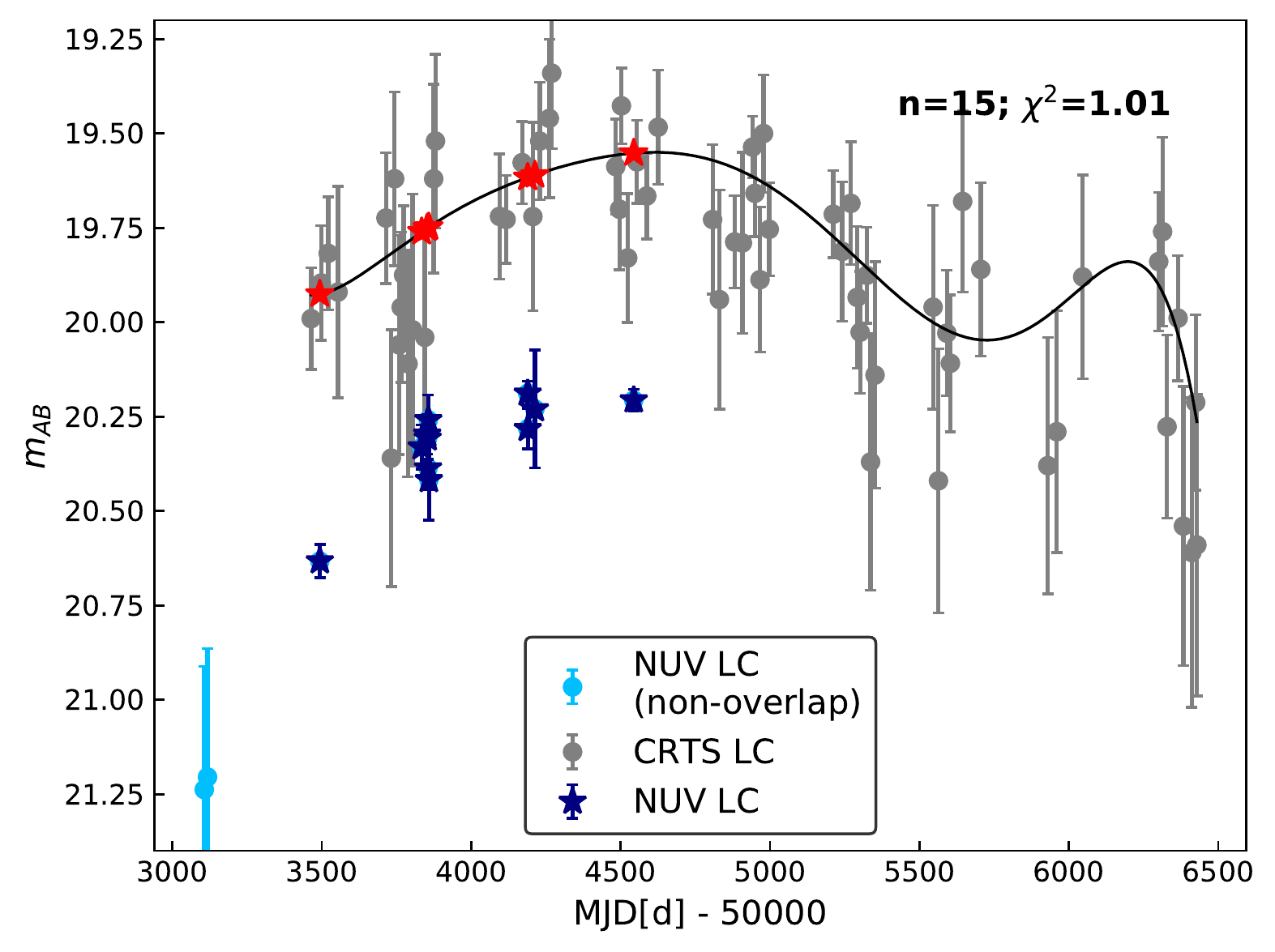}
 \caption{This example source, SDSS J125731.3+001454,
   has an optical light curve (LC; data shown by gray circles with error
   bars) that is well described by a best-fit polynomial of order $n=15$ with a reduced $\chi^2=1.01$ (black curve).
   The corresponding UV data points outside and inside
   the time interval that overlaps with the optical data are denoted by light
   blue circles and dark blue stars, respectively.  The red stars mark
   the optical data interpolated to the times of the UV observations,
   using the best-fit polynomial. The cross-correlation coefficient
   measured between the red and dark blue stars is $R=0.86$.}

 \label{fig:method-obs}
\end{figure}

Fig.~\ref{fig:method-obs} illustrates a typical example of a pair of optical and UV light curves, along with the best-fitting polynomial for the optical data (shown with the black line). We computed the cross-correlation coefficient, $R_i$, based on the overlapping part of the optical and UV light curves. Even though the polynomial fits are a reasonable fit for the optical light curves, they cannot be extrapolated beyond the range of available data. Therefore, we used only the UV data points whose dates fall inside the optical baseline (stars in dark blue in Fig.~\ref{fig:method-obs}) and excluded the UV points outside this range (light blue points in Fig.~\ref{fig:method-obs}). We then selected the points from the  polynomial fit that correspond to the times of the UV observations, giving us new optical points (red stars in Fig.~\ref{fig:method-obs}) that are simultaneous with the UV.

Having calculated the magnitudes at $N$
coincident times, the standard Pearson cross-correlation coefficient can then be straightforwardly evaluated

\begin{equation}  \label{eq:correlation}
    R_\mathbf{XY}^2 \equiv \frac{[Cov(\mathbf{X},\mathbf{Y})]^2}{[Var(\mathbf{X})][Var(\mathbf{Y})]},
\end{equation}
where $\mathbf{X}=(x_1,x_2,...,x_N)$ and $\mathbf{Y}=(y_1,y_2,...,y_N)$
represent the optical and UV data vectors, at times $(t_1,t_2,...,t_N)$.  
With the above method, we obtained the cross-correlation coefficient $R_i$ for each of the 1,315 sources and derived the
corresponding distribution of the correlation coefficient, $P(R_{\rm obs})$. 

\subsubsection{Modeling and Simulations} \label{sec:modeling-simulation}

We performed a theoretical study of the optical/UV variability of the population of quasars to explore whether the observed distribution $P(R_{\rm obs})$ reflects genuine correlations.

First, we conducted a null test, in which the optical and UV light curves are completely uncorrelated. For this, we randomly shuffled the pairings of the optical and UV data, i.e. every optical light curve is paired up with the UV light curve from a different quasar, rather than from itself, so that no correlation is expected. We then repeated the calculation of the $R_i$ for each pair of UV and optical light curves in this shuffled sample, following the steps detailed in \S~\ref{sec:observed_data} and obtained the distribution of $R_i$ for this null test $P(R_{\rm null})$. The advantage of this test is that does not make any assumption for the underlying variability, since it uses only observed data.

Next, we explored the level of correlations for the population by generating model light curves with properties (sampling, photometric errors) similar to the observed ones, and performing the same cross-correlation analysis as above. 

\begin{figure}
 \includegraphics[width=\columnwidth]{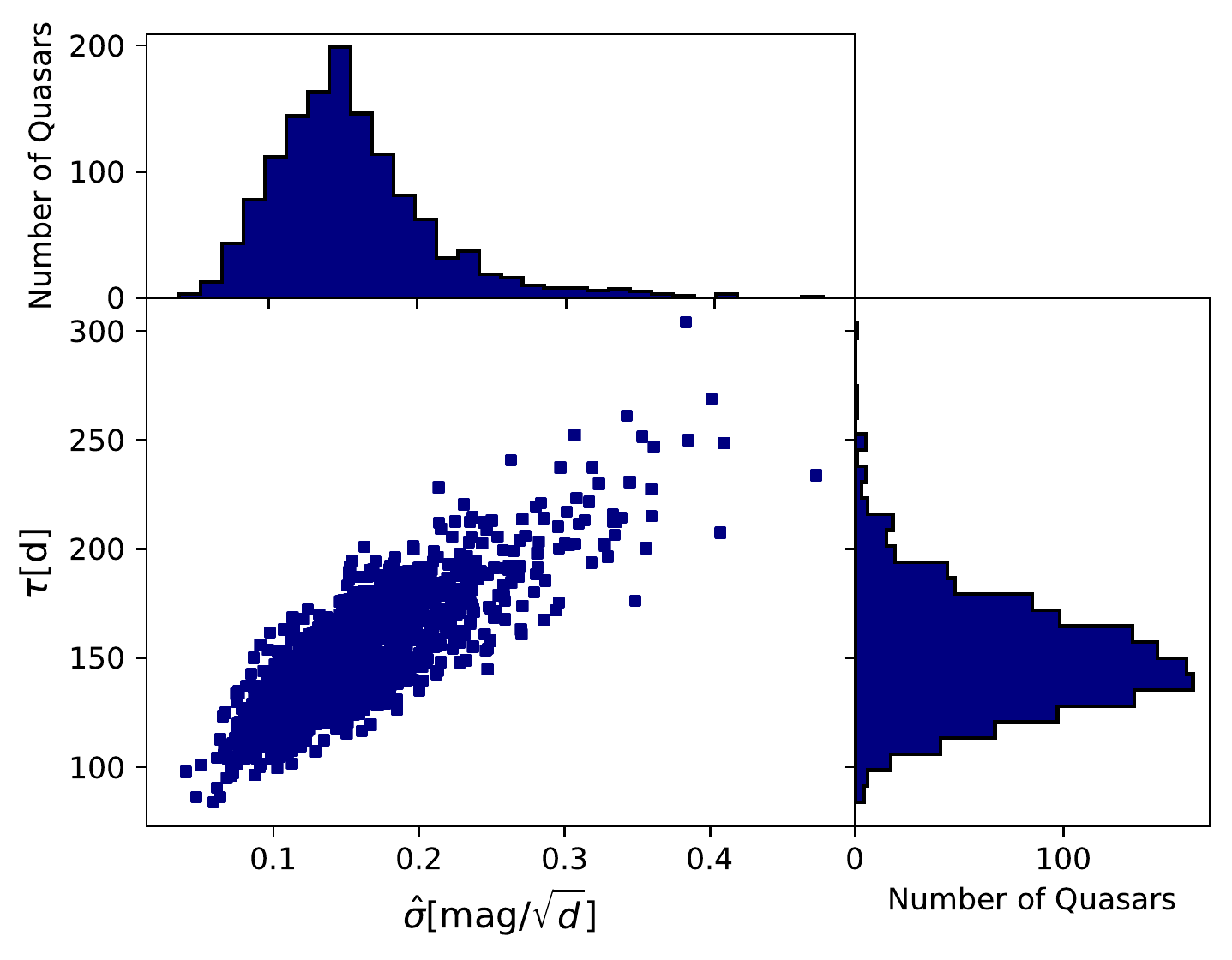}
 \caption{Damped random walk (DRW) model parameters for the sample of 1,315 quasars estimated from their properties adopting correlations from \citet{Macleod2010}.}
 \label{fig:best-fit-drw}
\end{figure}

In particular, we used the DRW model, which provides a successful description of quasar variability in optical and NUV \citep{Macleod2010,Zhu2016}. 
The DRW parameters (the characteristic correlation time-scale $\tau$, and the variability amplitude $\hat{\sigma}$) depend on global quasar properties, such as the redshift, the absolute i-band magnitude and the black hole mass. For each quasar, we estimated the DRW parameters, using eq.~7 and Table~1 from \citet{Macleod2010} for the $SDSS$ $r$-band.\footnote{The $SDSS$ $r$-band is the closest to Johnson~$V$ band, in which the {\it CRTS} data are callibrated.} 
For this, we extracted the redshift from the HMQ catalogue, and calculated the absolute i-band magnitude, and from this the BH mass, following the steps in \citet{Charisi2016}.  The exception is that for 373 quasars in our final sample, virial mass estimates were available, which we adopted from \citet{Shen2008}. 
This process produced 6 outliers with unrealistically large virial masses ($\sim 10^{13-14} M_{\odot}$), likely misclassified blazars,\footnote{The estimation of the BH mass from the i-band luminosity from \citet{Shen2008} and \citet{Macleod2010} does not hold for blazars.} which we excluded from the sample (see also \ref{sec:sample_selection}). The estimated DRW parameters are shown in Fig.~\ref{fig:best-fit-drw}.

With the DRW parameters estimated above, we generated continuous mock light curves, sampled at 1-day intervals. We examined two scenarios: (A) a perfect correlation between optical and UV variability. In this case, the UV light curve is a rescaled version of the optical, with the relative amplitude $r_{q}\equiv\sigma_{\rm opt}/\sigma_{\rm UV}$ assumed to be a free parameter.
(B) the optical and UV data are not
fully correlated. For this, we assumed that a fraction of quasars $f_{\rm cor}$ have fully correlated optical-UV variations, while the remaining ($1-f_{\rm cor}$) have uncorrelated variations in the two bands (see also \S~\ref{sec:partial-cor}). For the uncorrelated sources, we generated random realisation of the optical and UV light curves independently.

We down-sampled the continuous DRW light curves generated above, at the observational times by {\it CRTS} and {\it GALEX}, respectively. We added Gaussian deviates with zero mean and standard deviation equal to the photometric uncertainty of each point to incorporate the measurement errors.
By preserving the observed properties in the simulated light curves, we account for systematic effects introduced by the quality of the data. From the down-sampled optical and UV light curves, we calculated the cross-correlation, as before, by interpolating the optical light curve with an $n^{\rm th}$-order polynomial, finding the overlapping interval and selecting the times that correspond to the UV observations. We repeated the process for the entire population, and we obtained the distribution $P(R_{\rm model})$.

Our analysis has two free parameters, the relative amplitude of variability $r_q$, which we allowed to vary from $0.5$ to $4.5$ and the correlated fraction $f_{\rm cor}$, which we varied between $0$ and $100\%$. We sampled the two parameters on an equally spaced 20$\times$20-grid in this two dimensional parameter space. For each pair of ($f_{\rm cor},r_{\rm q}$), we generated 20 independent realisations of the population, resulting in 20 independent $P(R_{\rm model})$-distributions for each of the $20\times20=400$ pairs of ($f_{\rm cor},r_{\rm q}$). 

We compared the simulated realisations $P(R_{\rm model})$ with the observed $P(R_{\rm obs})$, using the Kolmogorov-Smirnov test (KS-test). In particular, we computed the KS distance ($D_{\rm KS}$) between the normalized cumulative distribution function (CDF) of $R_{\rm obs}$ and each
$R_{\rm model}$.
This provided 20 $D_{\rm KS}$ values for each pixel on the ($f_{\rm cor},r_{\rm q}$) plane. We used the average, $\overline{D_{\rm KS}}$, as the
figure-of-merit to determine which set of parameters better reproduces the observed distribution, with lower values corresponding to better fits to the data.

\section{Results} \label{sec:results}
\begin{figure}
 \includegraphics[width=\columnwidth]{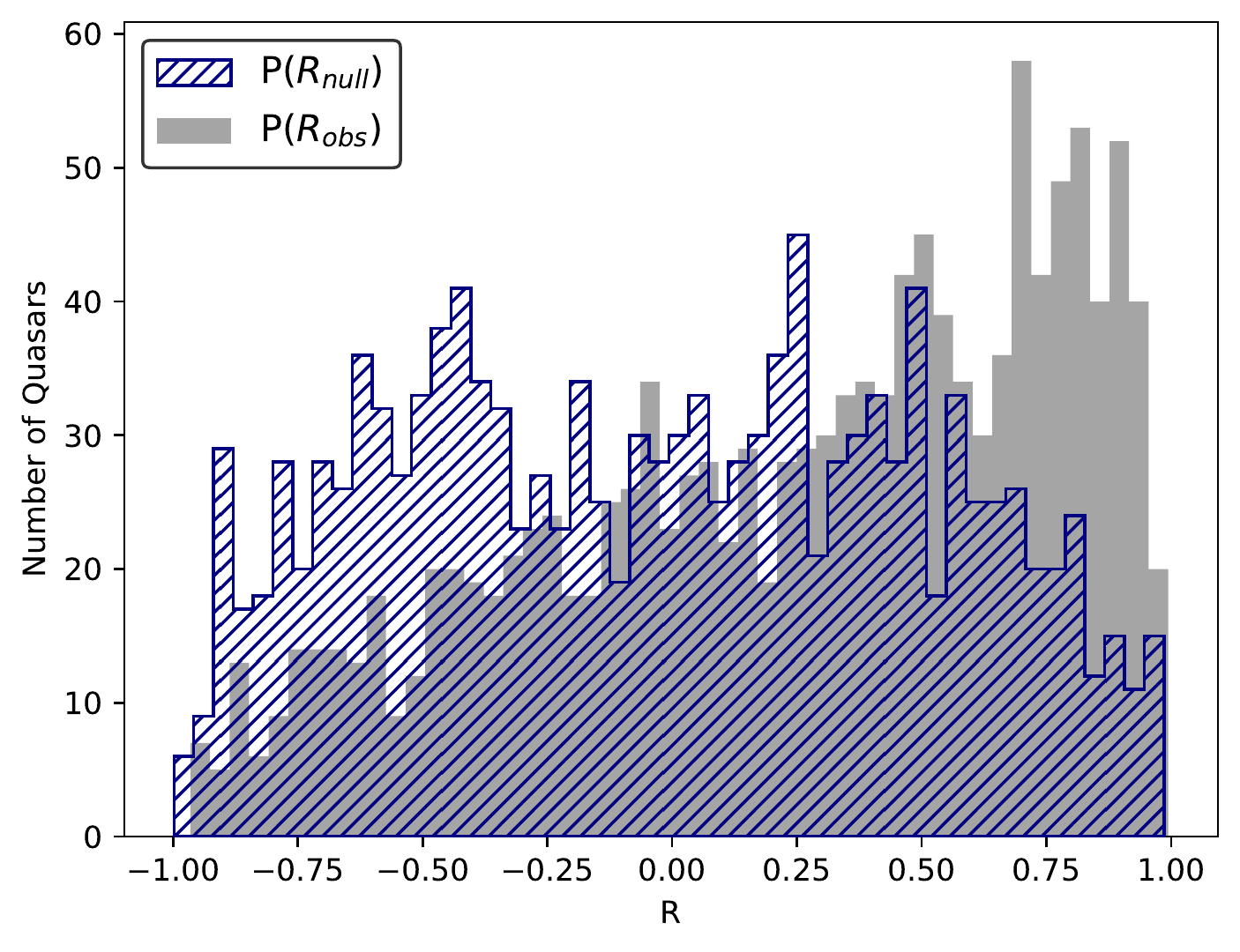}
 \caption{Distribution of the cross-correlation coefficient $R_{\rm obs}$ between the optical and UV variability in our sample of 1,315 quasars (gray). The hatched histogram shows the same distribution obtained from a null-test, in which the optical and UV light curve pairings were randomly shuffled, so that no correlations are expected.  The comparison between the two histograms reveals that the correlations in the data are genuine and are not produced by poor data quality alone.} 
 \label{fig:Robs_Rnull}
\end{figure}

We calculated the cross-correlation coefficient for the population of 1,315 quasars in our final sample. In Fig.~\ref{fig:Robs_Rnull}, we show the distribution of cross-correlation coefficients $R_i$ for the observed data $P(R_{\rm obs})$ with a solid gray histogram. The distribution shows a clear trend towards large $R_{\rm obs}$ values, with a broad peak at $0.75\lsim R_{\rm obs}\lsim 1$. 

We then performed a model independent null test, shuffling the optical and UV pairs of light curves. The distribution of $R_i$ for this null test $P(R_{\rm null})$ is shown by the blue hatched histogram in Fig.~\ref{fig:Robs_Rnull}. We see that in the limit of no correlation, the distribution $P(R_{\rm null})$ is relatively flat for all values of $R_i$. 
In the ideal case, a delta function at $R=0$ would be expected, but the sparse data flatten the distribution; we further explore the systematics introduced by the limited data quality in \S~\ref{sec:idealized-test-photErr}.

The excess of values at the positive correlation end in the observed distribution $P(R_{\rm obs})$ compared to the flat $R_{\rm null}$ distribution is a strong indication that the optical and UV variability is positively correlated in our sample.
In other words, although the sub-optimal quality of data (e.g., due to sparse sampling, significant photometric errors, non-simultaneous observations, etc.) may  smear the apparent $R$-distribution, they do not, by themselves account for the skewed observed shape in $P(R_{\rm obs})$.

\begin{figure*}
    \centering
    \includegraphics[width=\textwidth]{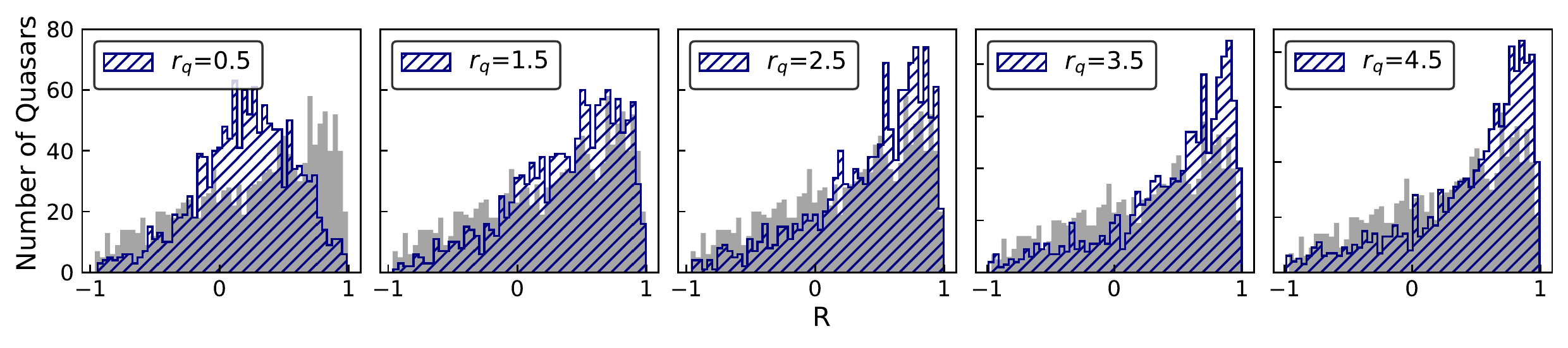}
    \caption{Distributions of the cross-correlation coefficient $P(R_{\rm model})$ assuming fully correlated UV and optical data (blue hatched histograms), for different amplitude ratios, $r_{\rm q}$, compared to the observed distribution (gray; same in every panel).}
    \label{fig:r-q-dist}
\end{figure*}

Next, we compared the observed distribution $P(R_{\rm obs})$ with theoretical distributions produced from simulated light curves $P(R_{\rm model})$. In Fig.~\ref{fig:r-q-dist}, we show model distributions, assuming that the optical and UV variability are fully correlated (scenario A), while varying the relative amplitude of variability $r_q$. We see that the simulated distributions begin to approach the observed one as the variability amplitude is increased, with $r_q$ between 1.5 and 2.5 providing the best fit.
For even larger $r_q$, the simulated histograms show overly skewed positive correlations.   This is expected: as the UV variability stands out more above the noise, the intrinsic correlations show up more accurately in the measurements. The best-fit values of $1.5\lsim r_q\lsim 2.5$ are consistent with previous studies, which showed that quasars tend to have higher variability amplitudes at shorter wavelengths \citep{Welsh2011,Gezari2013,Zhu2016,Charisi2018}.
Another important feature is that none of the model distributions for fully correlated variability can reproduce the observed distribution exactly; there is always an excess of $R_i$ at negative values ($R<0$). 

\begin{figure*}
    \centering
    \includegraphics[width=\textwidth]{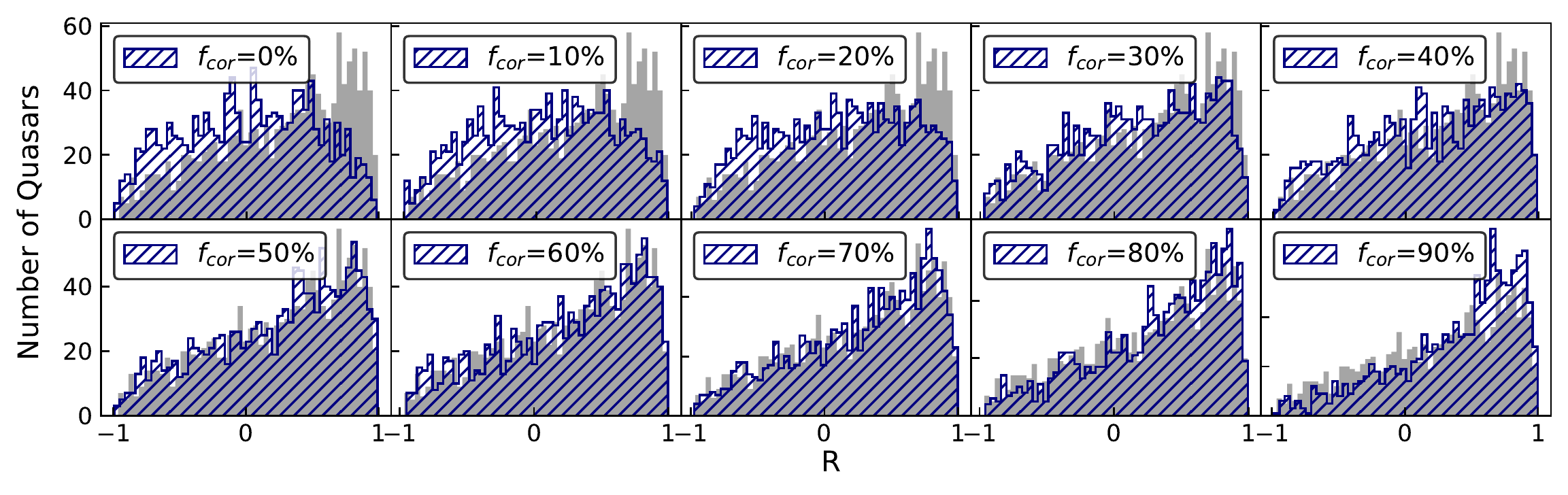}
    \caption{Distributions of the cross-correlation coefficient $R_{\rm model}$ predicted for different correlated fractions $0\leq f_{\rm cor}\leq 90\%$ at fixed variability amplitude ratio $r_{\rm q} = 2.5$ (blue hatched), compared to the observed distribution (gray; same in every panel).}
    \label{fig:f-cor-dist}
\end{figure*}

The above finding led us to explore scenario B, in which we vary the fraction of quasars that are fully correlated $f_{\rm cor}$ for a fixed variability amplitude. We chose $r_{\rm q} = 2.5$, since, as seen in Fig.~\ref{fig:r-q-dist}, this variability amplitude provides a relatively good fit to the observed distribution.
The resulting distributions are shown in Fig.~\ref{fig:f-cor-dist}.\footnote{We did not reproduce the case with $f_{\rm cor}=100\%$ in Fig.~\ref{fig:f-cor-dist} --- this is the same as the middle panel in Fig.~\ref{fig:r-q-dist}, where all quasars have fully correlated optical and UV variability.} 
As this figure shows, for $f_{\rm cor} \approx 60\%$, the simulated distribution $P(R_{\rm model})$ appears consistent with the observed $P(R_{\rm obs})$, whereas for higher/lower values of $f_{\rm cor}$, it becomes too steep or missing the peak near $R\approx 1$, respectively.

\begin{figure}
 \includegraphics[width=\columnwidth]{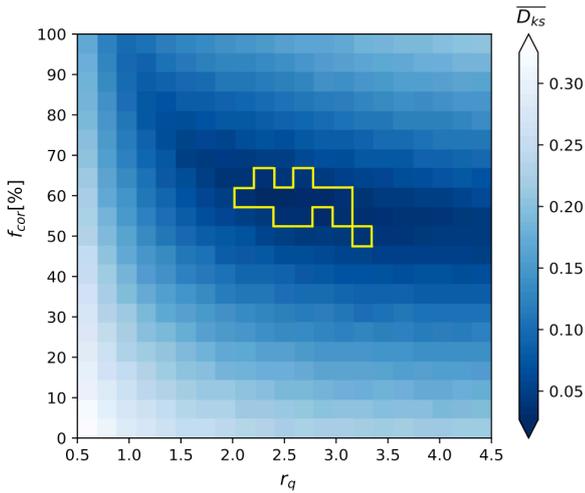}
 \caption{Average KS distance, $\overline{D_{\rm
       KS}}$, as a function of the fraction of fully correlated
   quasars, $f_{\rm cor}$, and amplitude ratio of UV-to-optical DRW
   variability, $r_{\rm q}$. The yellow contour marks an estimate of the
   95$\%$ confidence region.}
 \label{fig:result_contour}
\end{figure}

The above two figures demonstrate the need to fully explore the parameter space, considering both the fraction of quasars that are fully correlated $f_{\rm cor}$ and the amplitude ratio between stochastic UV and optical variability $r_{\rm q}$ as free parameters. For each pair of ($f_{\rm cor}$, $r_q$), we generated 20 random realisations of the population and calculated the KS distance for each.
In Fig.~\ref{fig:result_contour}, we present our figure-of-merit, i.e. the average (from 20 realisations)
KS distance between the modelled and the observed $R$-distributions ($\overline{D_{\rm KS}}$), over our 2D parameter space. The colour-bar on the right side of the figure
indicates different values of $\overline{D_{\rm KS}}$. The yellow contour on this 2D grid encloses all pixels which had the lowest $\overline{D_{\rm KS}}$ in any one of the 20 realisations. These represent the set of models which were found to be best-fit models (i.e. the combination of parameters that returned the lowest $\overline{D_{\rm KS}}$) in any of our 20
realisations.\footnote{There are fewer than 20 pixels inside this
  contour, because two or more realisations sometimes yield the same
  best-fit pixel.} 
We interpret this region as our $95\%$
confidence region (i.e. the chance that the best-fit model falls
outside this region is $\lsim5\%$).

\begin{figure}
 \includegraphics[width=\columnwidth]{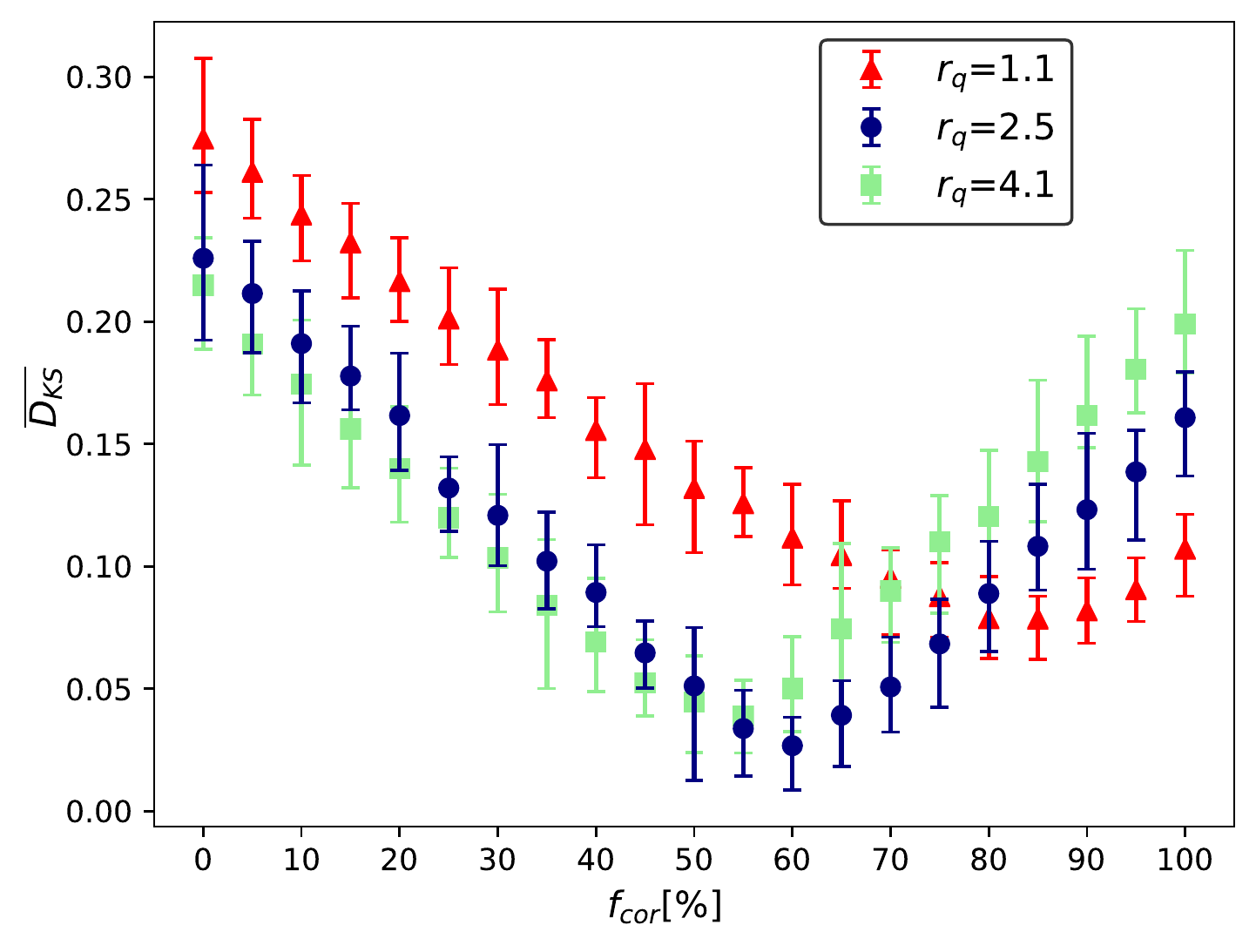}
 \caption{Average KS distance $\overline{D_{\rm KS}}$ and error bars (representing the full range of $D_{\rm KS}$ over 20 realisations) as a function of $f_{\rm cor}$ for $r_{\rm q}$ = 1.1, 2.5 and 4.1 (as labeled in the inset).}
 \label{fig:ave-KS-err}
\end{figure}

Since Fig.~\ref{fig:result_contour} focuses on the average KS distance and does not show the distribution of values from the individual realisations, in Fig.~\ref{fig:ave-KS-err}, we show the figure-of-merit $\overline{D_{\rm KS}}$ as a function of $f_{\rm cor}$ at fixed $r_{\rm q}$, with error bars that represent the full range of values we obtained from the 20 realisations. Red triangles, blue circles and green squares illustrate $r_q$= 1.1, 2.5 and 4.1, respectively.
The relatively small error bars demonstrate that the observed minima are real and the observed trends are not sensitive to individual random realisations of the population. As a result, the trend and peaks we see in Fig.~\ref{fig:result_contour} are robust.

Overall, the figure-of-merit based on the  KS-test shows a well-defined peak at
($f_{\rm corr},r_{\rm q}$)$\approx$($60 \%$, 2.5), representing
the best-fit parameter values, for which the predicted
$R$-distribution is closest to the one observed. This corresponds to a
simple model in which the UV {\it vs.} optical DRW amplitude ratio is
always 2.5, and in which $60\%$ of all quasars have fully correlated
UV {\it vs.} optical variability (with $R=1$, otherwise $R=0$). This
model provides a good match to the observed $R$-distributions.

A corollary of our finding is that the simplest scenario, in which all
quasars have well-defined correlations between optical and UV
variability, as may have been inferred from prior studies on a small 
sample of $\sim$ two dozen AGN, is ruled out by our analysis at high confidence.
We note however that in our analysis we simulated a population with a fixed constant variability amplitude $r_q$. It is possible that if we varied $r_q$, we could reproduce the observed distribution $R_{\rm obs}$, even for fully correlated light curves. We will explore this case in future work.

\section{Discussion} \label{sec:discussion}

\subsection{Sampling and photometric errors} 
\label{sec:idealized-test-photErr}
The null test of completely uncorrelated optical and UV variability returned a flat distribution $R_{\rm null}$. Ideally, we should infer a distribution similar to a delta function, with a peak at $R=0$. This means that the limited data quality can introduce spurious cross-correlations ($R \neq 0$). This could potentially be improved, if the light curves were sampled at higher cadence, had smaller photometric errors, or longer baselines. Here we examine the impact of these three aspects.

\begin{figure}
 \includegraphics[width=\columnwidth]{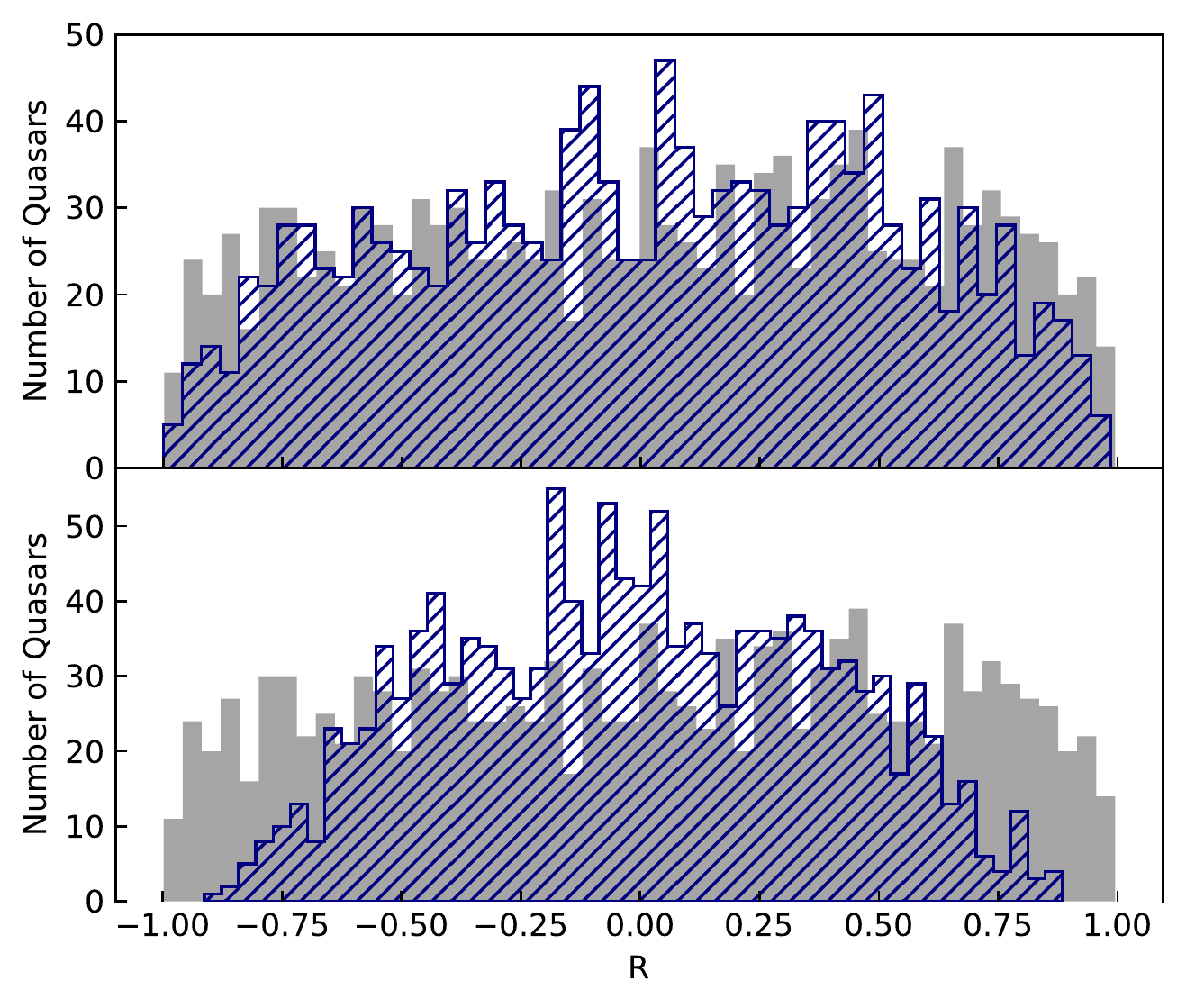}
 \caption{Distributions of the cross-correlation coefficients (solid gray; same in both panels) of mock optical and UV datasets, assuming uncorrelated DRW variability in the two bands (true $R=0$), noise ratio $r_q=2$, and no photometric errors.  The dark blue hatched histogram in top panel denotes the distribution obtained from 10 times more densely sampled UV time-series, whereas the hatched histogram in the bottom shows the distribution with the photometric errors included (same as the last panel with $f_{\rm cor}=0$ in Fig.~\ref{fig:f-cor-dist}).}
 \label{fig:uv-sampling}
\end{figure}

First, we explored systematics from the limited quality of the UV light curves, generating simulations of uncorrelated light curves ($R=0$), with a fixed variability amplitude ratio at $r_q=2.5$ (as in Fig.~\ref{fig:f-cor-dist}). In order to understand the effect of photometric errors of the UV simulated light curves, we set the photometric errors to zero (practically, we generated noiseless DRW light curves), and kept the original cadence and baseline of the observations. The resulting distribution is shown by the gray
histogram in the top panel of Fig.~\ref{fig:uv-sampling}.
For reference, we also show the distribution, in which the photometric errors were included in the simulations in blue; this is the same as the top left panel of Fig.~\ref{fig:f-cor-dist}. The distributions (from simulations with and without UV photometric errors) are very similar, which indicates that the photometric errors have a negligible effect and are not the main culprit for flattening the $R$-distribution. The only visible differences appear near $R\approx \pm1$: the removal of purely random noise in the UV time-series makes it less probable for such strong (anti)correlations to arise by chance.

Then, we explored the effect of sampling, generating high-cadence mock UV light curves. For this, we inserted 9 additional data points, spaced evenly between each two consecutive observed times, such that the resulting UV MJDs are 10 times more densely sampled. We down-sampled the mock DRW light curves at the new UV MJDs to obtain UV light curves with higher cadence and no photometric errors. This new sample yields the blue hatched histogram 
at the bottom panel of Fig.~\ref{fig:uv-sampling}. As the figure shows, the gains are relatively modest, and the expected delta-function at $R=0$ remains highly smeared out, despite the 10-fold increase in the density of the UV time-series.

\begin{figure}
 \includegraphics[width=\columnwidth]{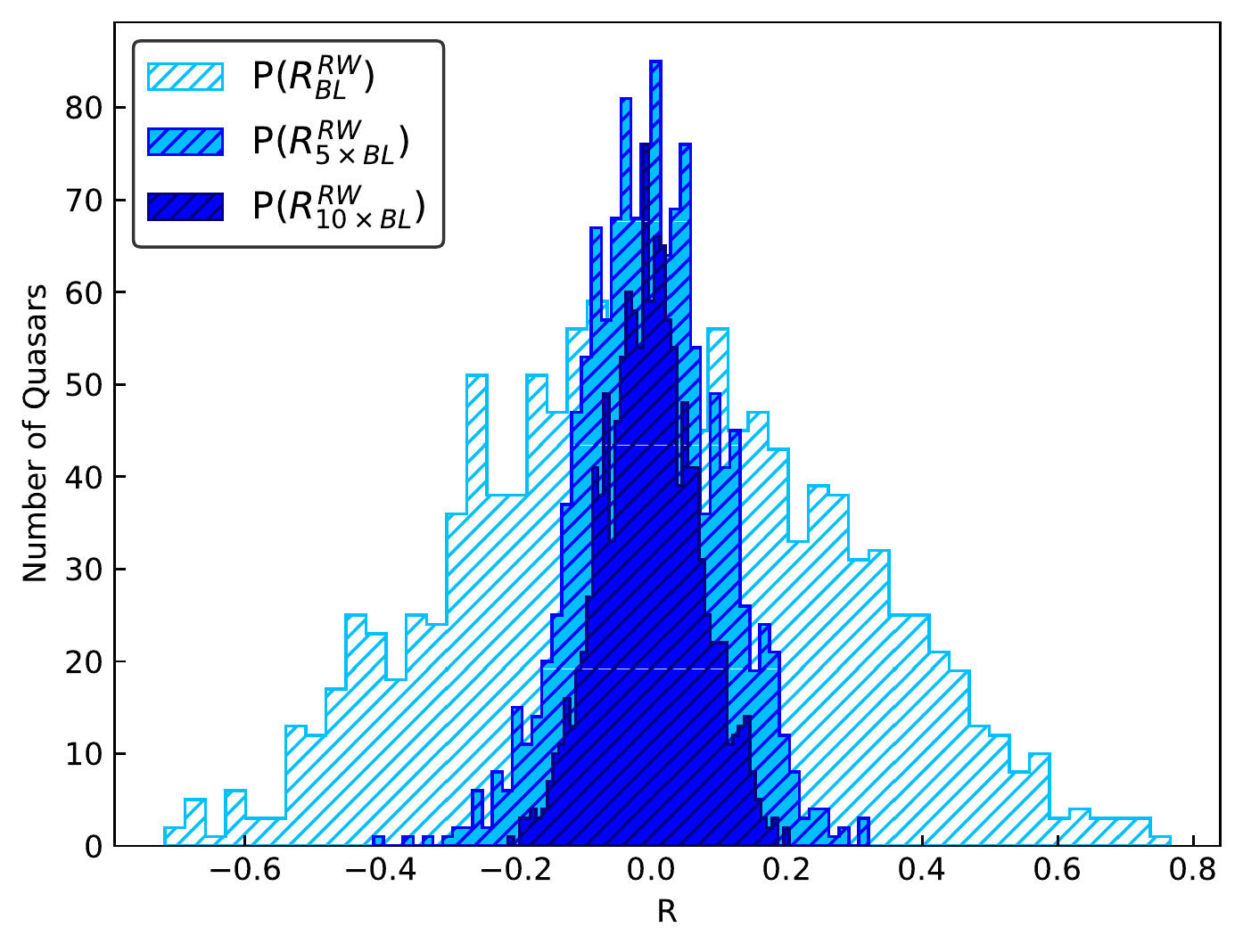}
 \caption{Probability distributions of the cross-correlation coefficients of DRW optical and UV datasets with different baselines, assuming uncorrelated DRW variability in the two bands (true $R=0$) and noise ratio $r_q=2.5$. The light blue histogram denotes the distribution obtained from DRW datasets with the same baselines as the {\it CRTS} light curves. The medium and dark blue histograms show distributions where the baselines in DRW datasets are 5 and 10 times longer, respectively.}
 \label{fig:rw}
\end{figure}

Finally, we explored a more idealized case.
We generated optical and UV DRW time-series with 1 day cadence, and with the same baselines (i.e. the baselines of the optical DRW light curves) in the two bands.
We repeated these idealized simulations with extended baselines of either 5 or 10 times their original optical baselines.
The three distributions are shown in Fig.~\ref{fig:rw}, where $\rm P(R^{\rm RW}_{\rm BL})$ (light blue) shows the
case, where the length of the DRW light curves is fixed at the observed value of the initial optical light curves
plus $P(R^{\rm RW}_{\rm 5 \times BL})$ (medium) and $P(R^{\rm RW}_{\rm 10 \times BL})$ (dark) show the distributions for the extended light curves with 5$\times$ and 10$\times$ longer baselines, respectively.
The light blue distribution shows a significant improvement compared to the dark blue distribution in Fig.\ref{fig:uv-sampling}, which means that the even denser and coincident sampling both for the optical and UV light curves reduces the chance of detecting correlations by chance. 
However, the distribution is still relatively broad.
As can be seen from Fig.~\ref{fig:rw}, the distributions obtained from DRWs with longer baselines converge more closely to the expected delta function. 

\subsection{Partial Correlations} \label{sec:partial-cor}

We tested a simple scenario, in which a fraction of quasars are fully correlated and the rest have fully uncorrelated UV and optical variability. In general, there could be different levels of partial optical-UV correlations, and this level can be different from one source to another, i.e. each quasar could have a partially correlated optical-UV variability with $R \in [0,1]$. In fact, this already seems to be the case for the well-sampled light curves in \citet{Edelson2019}. We calculated the cross-correlation coefficients for these sources between the $V$ and M2 bands\footnote{The M2 band is the closest to {\it GALEX} NUV.} light curves (NGC 4151, $R=0.81$; NGC 5548, $R=0.92$; NGC 4593, $R=0.62$; Mrk 509, $R=0.97$). 
Note that, since these light curves are well-sampled, for this calculation we omitted the polynomial interpolation of the optical light curve, which likely introduces additional biases.

In this paper, we focused on testing the first hypothesis (a fraction of quasars have fully uncorrelated data), which is the simplest to implement. However, given the quality of the data, we expect that the two cases will likely produce indistinguishable distributions. While we will not evaluate the second case here, we defer to future work to assess how well a population of quasars with partially correlated light curves fits the same data analysed here. This can be attained by generating pairs of mock DRW light curves whose input is $R<1$.   

\subsection{Time delays} \label{sec:time-delay}

If the UV/optical luminosity arises via reverberation of a central illuminating source, the variability should track one another, but with a time-delay that corresponds to their relative distances to the central source. Typical delays are expected to be on the order of days (see, e.g., \citealt{Edelson2019} and references therein).
Even for the massive quasars in our sample, with SMBH mass of $\sim 10^9$ $M_{\odot}$ the light-crossing time is $\sim$12 days, given the expected size of the optical emission, which is $\sim$100 $R_{\rm S}$, where $R_{\rm S}$ is the Schwarzschild radius. Observationally, the size of the optical emission region is smaller (e.g., Fig.~1 in \citealt{Morgan2007} shows
$< 10^{16}$ cm, corresponding to a light-crossing time of $\sim$4 days).

Our data are not sufficiently well sampled to be sensitive to these short delays. We verified this expectation by simulating optical and UV light curves with time lags between 1-100 days and repeating our analysis. For the simulations here, we adopted $f_{\rm cor} = 60\%$ and $r_{\rm q} = 2.5$. We calculated the average KS distance and found no significant difference.  We concluded that more densely sampled light curves are necessary to probe time-lags expected in reverberation models of thin disks.

\subsection{KS Probability} \label{sec:p-ks}
\begin{figure}
 \includegraphics[width=\columnwidth]{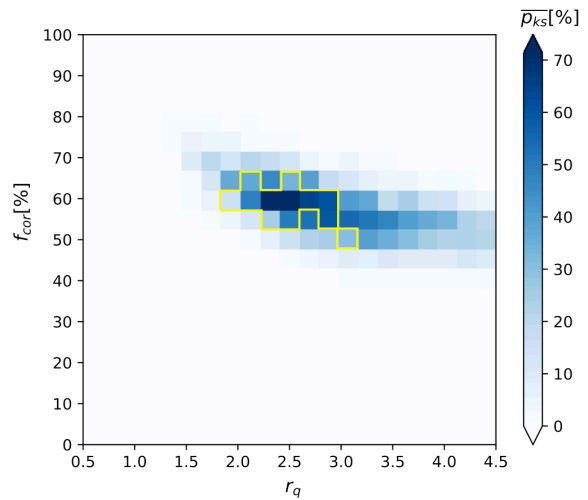}
 \caption{Average KS probability from 20 realisations for each pixel as a function of the fraction of quasars $f_{\rm cor}$ and the variability amplitude $r_{\rm q}$ The yellow contour indicates the 95$\%$ confidence level (same as the yellow contour in Fig.~\ref{fig:result_contour})}
 \label{fig:p-ks}
\end{figure}

In our analysis above, we quantified how well a model distribution fits the observed data using the KS distance. In the standard version of the KS test, the KS distance corresponds to a probability that the examined distribution is drawn from a reference distribution. In our case, this can be loosely translated into the likelihood that our distribution $R_{\rm obs}$ was drawn from the predicted probability distribution $R_{\rm model}$. 
However, the results need to be interpreted with caution, since the assumptions do not obey the formal requirements to convert the KS distance to probability. In particular, the $R_i$ values are not drawn independently from the same distribution, because each $R_i$ is calculated for a different quasar. Since the properties of the individual pairs of light curves are not dramatically different, the obtained probabilities should be approximately correct.
With this caveat in mind, for each realisation of the population presented in Fig.~\ref{fig:result_contour}, we calculated the KS probability. In Fig.~\ref{fig:p-ks}, we show the average probability ($\overline{p_{\rm KS}}$) 
for each of the 20$\times$20 pixels on the ($f_{\rm cor},r_{\rm q}$) plane. We see that this figure delineates a similar region as the 95\% contour we identified in Fig.~\ref{fig:result_contour}, providing additional justification for identifying this as our 95\% confidence region.

\section{Conclusions}
\label{sec:conclusion}

In this paper, we performed a study of the correlation between UV and optical variability of quasars. These inter-band correlations are a useful probe of the physics of accretion disks in quasars, but they are challenging to measure, mostly because time-domain data in the UV are sparse. As a result, previous studies focused on a handful of quasars with targeted observations and well-sampled light curves in both bands; they found strongly correlated variability. Here we complemented previous studies and instead assembled a large sample of 1,315 quasars, selected from {\it GALEX} and {\it CRTS}.  While this necessitated a compromise in the data quality for the individual light curves, we were able to analyse this sample statistically, and extract the intrinsic correlations between the optical and UV light curves for the population as a whole. Our analysis utilized mock light curves, which mimic the important features of the data (such as sampling, baseline, and photometric errors). We also performed a model-independent null test, in which we random shuffled the pairings of the optical {\it vs.} UV light curves.

We found that strong correlations exist in this much larger sample, but we ruled out, at $\sim$95\% confidence, the simple hypothesis that the intrinsic UV and optical variations of all quasars are fully correlated. We explored a simple model, in which a fraction $f_{\rm cor}$ of quasars have UV light curves that are fully correlated with the optical, but with an amplitude that is scaled by a factor of $r_q$, while the remaining fraction  $(1-f_{\rm cor})$ are not correlated at all. We found that the values of $f_{\rm cor}\approx 60\%$
and $r_q\approx 2.5$ best reproduce the observed distribution of correlation coefficients. Therefore, our results imply the existence of physical mechanism(s) that can generate uncorrelated optical and UV flux variations, such as expected, for example, from local temperature fluctuations~\citep{Dexter2011}.

Future work should extend our analysis, by improved modeling of intrinsic quasar variability, and allowing for partially correlated UV {\it vs.} optical fluctuations.  Our study was also limited by the available UV data, ultimately resulting in a cut from $\sim150,000$ quasars down to a sample of $\sim 1,000$ with sufficient overlapping optical and UV data. A future UV time-domain survey would yield great improvements. Our analysis in \S~\ref{sec:discussion}, especially Fig.~\ref{fig:rw}, further indicates that apart from the number of sources, the main limitation for our study was the length of the baseline: the best improvement would be provided by a large-area UV and optical surveys with simultaneous observations covering at least several years.

\section*{Acknowledgements}
We thank Suvi Gezari for helpful discussions, and Rick Edelson for providing the {\it Swift} light curves.  We acknowledge support from NASA through ADAP grant 80NSSC18K1093 (to ZH and DS) and through {\it Swift} grant 80NSSC19K0149 (ZH and MC), and from the National Science Foundation (NSF) through AST grant 1715661 (ZH) and through the NANOGrav Physics Frontier Center, award number 1430284 (MC). 

\bibliographystyle{mnras}

\bsp
\label{lastpage}
\end{document}